\documentclass{article}

\usepackage{amssymb,stmaryrd,epsfig}

\begin{document}

\title{The physical Church thesis and the sensitivity to initial conditions}

\author{Gilles Dowek}
\date{}
\maketitle
\thispagestyle{empty}

The {\em physical Church thesis} is a thesis about nature that
expresses that all that can be computed by a physical system ---~a
machine~--- is computable in the sense of computability theory.
At a first look, this thesis seems contradictory with the existence, in
nature, of chaotic dynamical systems, that is systems whose evolution
cannot be ``computed'' because of their sensitivity to initial
conditions.

The goal of this note is to show that there exist dynamical systems
that are both computable and chaotic, and thus that the existence in
nature of chaotic dynamical system is not, {\em per se}, a refutation
of the physical Church thesis. Thus, chaos seems to be compatible with 
computability, in the same way as it is compatible with determinism.

\section{The notion of computable function}

\subsection{Computability over the natural numbers}

Several equivalent definitions of the notion of computable function
may be given. One of the simplest is to define inductively a family 
$(C_p)_p$ where $C_p$ is a set of partial functions from 
${\mathbb N}^p$ to ${\mathbb N}$ by the six following rules 

\begin{itemize}
\item the functions $x_1, ..., x_p \mapsto x_i$ are elements of $C_p$, 
\item the function $x_1, ..., x_p \mapsto 0$ is an element of $C_p$, 
\item the function $x \mapsto x + 1$ is an element of $C_1$, 
\item if $f$ is an element of $C_q$ and $g_1, ..., g_q$ are elements of
$C_p$ then the composition of $f$ and $g_1, ..., g_q$, that is the
function $x_1, ..., x_p \mapsto f(g_1(x_1, ..., x_p), ..., g_q(x_1,
..., x_p))$ is an element of $C_p$, 
\item if $f$ is an element of $C_p$ and $g$ an element of
$C_{p+2}$ then the function $h$ defined by induction as follows 
$$h(x_1, ..., x_p, 0) = f(x_1, ..., x_p)$$
$$h(x_1, ..., x_p, y + 1) = g(x_1, ..., x_p, y, h(x_1, ..., x_n,y))$$
is an element of $C_{p+1}$,
\item if $f$ is an element of $C_{p+1}$ then the function $g$ defined by 
$g(x_1, ...x_n) = y$ if $f(x_1, ..., x_n, y) = 0$ and for all $z < y$ 
$f(x_1, ..., x_n, z) \neq 0$ is an element of $C_p$. 
\end{itemize}
In this last case, the function $f$ is not defined in $x_1, ...,
x_p$ if the function $z \mapsto f(x_1, ..., x_n, z)$ never takes the
value $0$.

\subsection{Computability on rational numbers}

The notion of computability extends easily to rational numbers.
Let $\alpha$ be the injection from ${\mathbb N}^2$ to ${\mathbb N}$
defined by $\alpha(n,p) = (n + p)(n + p + 1)/2 + p$. We define an
injection $\pi$ from ${\mathbb Q}$ to ${\mathbb N}$ by $\pi(r) =
\alpha(\alpha(sgn(r),num(r)),den(r))$ where $sgn(r)$ is the sign of 
$r$ ---~$0$ or $1$~---, $num(r)$ is the numerator of its reduced form
and $den(r)$ its denominator.  A partial function $f$ from ${\mathbb
Q}^p$ to ${\mathbb Q}$ is said to be computable if the partial function
from ${\mathbb N}^p$ to ${\mathbb N}$ mapping $\pi(x_1), ..., \pi(x_n)$
to $\pi(f(x_1, ..., x_n))$ is computable.

At a first look, this definition depends of the chosen injection $\pi$
from ${\mathbb Q}$ to ${\mathbb N}$. In fact, it can easily proved
that this is not the case.  If $\pi'$ is another injection from
${\mathbb Q}$ to ${\mathbb N}$ such that the function that map
$\pi'(x)$ and $\pi'(y)$ to $\pi'(x+y)$ and that that maps $\pi'(x)$
and $\pi'(y)$ to $\pi'(x y)$ are computable, then the set of
computable functions from ${\mathbb Q}^p$ to ${\mathbb Q}$ defined
with the injection $\pi'$ is the same as that defined with the
injection $\pi$.

\subsection{Computability on real numbers}

The extension of the notion of computability to real numbers is less
obvious, because the set of real numbers, unlike that of rational
numbers, is not countable.  A possible definition, that is equivalent
to the classical definition of Grzegorczyk and Lacombe
\cite{PourElRichards}, is that a function $g$ from an interval $I$ of
${\mathbb R}$ to ${\mathbb R}$ is computable if there exists a
computable function $G$ from ${\mathbb Q} \times {\mathbb Q}$ to
${\mathbb Q}$ and a computable function $\eta$ from the set of
strictly positive rational numbers
to itself, such that
$$|x - q| \leq \eta(\varepsilon)  \Rightarrow |g(x) - G(\varepsilon,q)| \leq
\varepsilon$$ 
In other words, to compute the value of $g(x)$ we must first chose a 
accuracy $\varepsilon$ with which we want to know the value of
$g(x)$. We must then apply the function 
$\eta$ to $\varepsilon$ to get the accuracy with which we are required
to supply the argument $x$. Then, we apply the function $G$ to
$\varepsilon$ and to a rational approximation $q$ of $x$ with an 
accuracy $\eta(\varepsilon)$ and we get the desired result. 
Repeating this  process, we can obtain 
approximations of the value of $g(x)$ as accurate as we want,
provided we can supply arbitrarily accurate approximations of 
$x$.

We can notice that a computable function from an interval 
$I$ to ${\mathbb R}$ is uniformly continuous on $I$.

\subsection{Non deterministic algorithms}

Let us finally mention another extension of the notion of computable
function that we will need. A non necessarily functional binary
relation $R$ is said to be computable if the function that maps $x$ to
the set of $y$'s such that $x~R~y$ is computable. This requires to
define a notion of computable function taking values in a set of sets.
Several solutions exist, depending on the cardinals of these sets 
and on the nature of their elements.

\section{The physical Church thesis}

Stating the physical Church thesis requires to consider a dynamical 
physical system, that is a system whose state evolves in time.
Time can be discrete or continuous. The state of the system can be
described by one or more discrete or continuous variables. It may
be directly accessible to measurement or not. It may evolve in a 
deterministic way or not.

We then consider the evolution of this system during a time interval
that starts at a predefined date and finishes either at another
predefined date or when the system satisfies, for the first time, a 
given property, that is itself computable ---~for instance, when 
one of its discrete variables describing its state takes for the first
time a given value.

We consider the evolution of the system during various time intervals 
defined as above and for which the initial states are different.

The physical Church thesis states that, whatever the considered
physical system is, the function ---~when the system is
deterministic~--- or the relation ---~when it is not~--- that relates
the initial state to the final state of the system is computable.

\section{An example of chaotic and computable system}

The existence, in nature, of chaotic dynamical systems,
that is system whose evolution can not be ``computed'' because 
of its sensitivity to initial conditions seems to contradict the
physical Church thesis.

In fact, as we shall see, there exist dynamical systems that are both 
chaotic and computable.  To do so, we shall give a very simple example
of such a system: {\em the baker's transformation}. It is a dynamical
system with a discrete time, in which a point of coordinate $x$ in
$[0,1]$ moves in one time step to the coordinate 
$b(x)$ where $b$ is the function defined by
$$b(x) = 2x~\mbox{if $x \leq 1/2$}$$
$$b(x) = 2 - 2x~\mbox{if $x > 1/2$}$$ 
\begin{center}
\epsfig{file=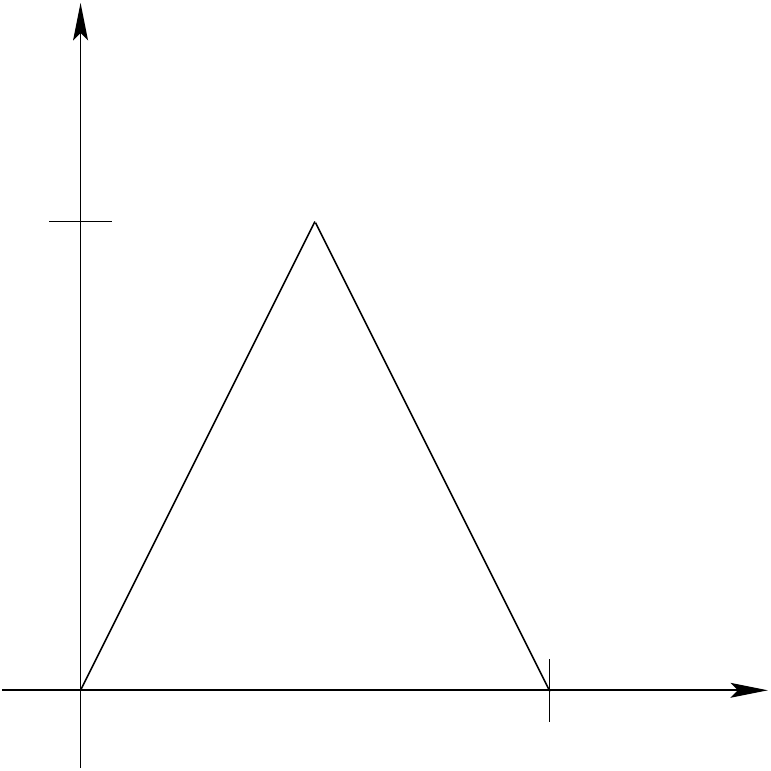,height=2cm}
\end{center}
if the initial state of the system is $x_0$, its state after $n$
time units is thus $b^n(x_0)$. 

It is not difficult to prove that this  system is chaotic: for all
$\eta$, if two initial positions have a difference less than 
$\eta$, after some time, the positions may differ of an arbitrary
value. 
Let $\eta$ be an arbitrary strictly positive real number and $a$ and
$a'$ two arbitrary real numbers in $[0,1]$. Let $n$ be an natural
number such that $1/2^n \leq \eta$.  The initial conditions $x_0 =
a/2^n$ and $x'_0 = a'/2^n$ are such that $|x_0 - x'_0| \leq
\eta$. However, after $n$ time units, $b^{n}(x_0) = a$ and $b^{n}(x'_0) =
a'$.

A very small difference in the initial conditions can thus change 
completely the state of the system after a certain time and knowing 
an approximation of the initial position, by a measurement for
instance, does not allow to predict the evolution of the system, even
approximately.

It is nevertheless elementary to prove that the function that maps an
initial position $x_0$ of the physical point and a date $n$ to the
position of the physical point at the date $n$, $F = x_0, n \mapsto
b^n(x_0)$, is a computable function from $[0,1] \times {\mathbb N}$ to
$[0,1]$.

What makes this system non predictable is not the fact that the
function $F$ is not computable, but the fact that to have a
reasonable approximation, let us say with an accuracy of $0.001$, of
the state of the system at a date $n$, it would be necessary to have
an approximation of the initial position with an accuracy $0.001/ 2^n$
and that for large enough values of $n$, it becomes impossible to know this
initial value by a measurement. It is therefore not the absence of
algorithm that makes this system non predictable, but the absence of
argument to supply to this algorithm.

It can be noticed that the construction of such an algorithm to which
no argument can be supplied is a common practice.  All the consistency
proofs of a theory $T$ (for instance, hyperbolic geometry) relatively
to that of a theory $U$ (for instance, euclidean geometry) that
proceed by constructing an algorithm that would transform a proof of a
contradiction in the theory $T$ into a proof of a contradiction in the
theory $U$, use such an algorithm.

The fact that no argument can be supplied to an algorithm does not 
mean that this algorithm does not exist.

\section{Two alternative descriptions}

\subsection{Really discrete positions}

The classical description of the baker's transformation describes the
position of a physical point in motion by a real number. This can be
criticized, assuming that, in reality, the position of a
physical point can take only a finite number of values in the interval
$[0,1]$ as it is the case in quantum physic, in some situations, for
instance when the state space has a finite dimension, and as it is
also the case, if we assume, like, for instance, Gandy \cite{Gandy},
that information has a finite density in nature, that is that a system
of finite size can have only a finite number of possible states.

Under such a description, the baker's transformation is not 
sensitive to initial conditions, because if we call $\eta$ the half of
the minimal distance that separates two distinct possible states, the 
condition $|x_0 - x'_0| \leq \eta$ is equivalent to $x_0 = x'_0$.

In this case, the function $b$ that maps the state of the dynamical
system at a given time, to its state at the next time step 
is a function of finite domain and thus it is computable. 
The function that maps the initial position 
$x_0$ of the physical point and the date $n$ to 
the position of the physical point at the date $n$, 
$F = x_0, n \mapsto b^n(x_0)$, is thus also computable.

\subsection{Discretely perceived positions}

Another way to criticize this description of the position of a
physical point by a real number is to accept to give a meaning only to 
effectively measurable values. This leads to reject this notion of
ideal position, that is not accessible to measurement, as
metaphysical.

In this case, the baker's transformation becomes a non deterministic 
dynamical system, as the sate of the system at a given time does not
determine its state at the next step. For instance, if we use to
measure the state of the system a device that provides three digits,
the initial state $0.000$ can be followed by $0.000$
or by $0.001$.  As above, the system is not sensitive to initial
conditions anymore and it is the fact that it is not deterministic
---~and not the fact that it is sensitive to initial conditions~---
that makes it non predictable.

In this case, as above, the number of possible measured values is 
finite and therefore the relation $R$ that relates two
measured values $x$ and $y$ if $y$ is one of the 
measured values that may follow $x$ is computable, because it is
finite.  The relation that relates a natural number $n$ and two values
$x$ and $y$ if $y$ may follow $x$ after $n$ steps is thus also computable.

\section{The limit state}

In the case of the baker's transformation, the sequence $(F(x_0,n))_n$
is not always convergent. Thus, we cannot speak, in general, of the
limit state of the system. In contrast, this becomes possible when we
consider dissipative dynamical systems. We can, in this case, discuss
the computability not only of the function $F$ that maps an initial
state $x_0$ of the system and a date $n$ to the state of the system at
this date, but also the function that maps an initial state $x_0$ of
the system to its limit state, that is to the limit of the sequence
$(F(x_0,n))_n$ when $n$ goes to infinity. In this case, it is not
difficult to find examples where this function is not computable. Any
example where this function is not continuous works.  Other examples
of non computable properties in chaotic systems are given, for
instance, in \cite{Hoyrup}.

This observation however is not in contradiction with the physical
Church thesis that is only concerned with the state of the system 
at a finite date $n$ and says nothing about its limit state when $n$
goes to infinity.

\medskip

The fact that a dynamical system may be both computable and non
predictable shows, once more, that the notion of computability largely
exceeds the practical notion of accessibility to computation.

\section*{Acknowledgments}

To Olivier Bournez, Jean-Baptiste Joinet, Giuseppe Longo and
Thierry Paul.

\end{document}